# Proposal for applying the novel gas-dynamic ion-beam extraction and bunching technique to the cryogenic stopping cells at FAIR

**Victor Varentsov**
*Facility for Antiproton and Ion Research in Europe (FAIR). Planckstraße 1. 64291 Darmstadt. Germany.*
*E-mail*: Victor.Varentsov@fair-center.eu

ABSTRACT: To enhance the quality of pulsed cold ion beams extracted from the two cryogenic stopping cells at FAIR (i.e., the one currently used in the FRS and the one under development for the Super-FRS at FAIR), we propose using a novel gas-dynamic ion beam extraction and bunching technique as an alternative to the radiofrequency quadrupole (RFQ) method. This technique allows for 100% ion transmission by placing a short stack of thin cylindrical electrodes behind the extraction RF carpet. Detailed gas-dynamic and ion trajectory computer simulations demonstrate that implementing this proposal will enable the achievement of world-record emittance values for ion beams in a wide mass range. The results of these simulations are presented and discussed.

KEYWORDS: RF buncher; RF carpet; cryogenic stopping cells at FAIR; gas-dynamic and Monte Carlo computer simulation, pulsed ion beam, trapping time.

## Contents

## 1. Introduction

A new cryogenic stopping cell (CSC) is currently under construction at the FAIR (Facility for Antiproton and Ion Research) for the purpose of conducting precise experiments with low-energy radioactive ion beams at the super-conducting fragment separator, Super-FRS. For example, this CSC is required for the Laspec (Laser Spectroscopy of short-lived nuclei) and MATS (precision Measurements of very short-lived of nuclei using an Advanced Trapping System for highly charged ions) experiments.

The prototype of the given CSC is currently employed in the projectile fragment separator FRS at GSI. A thorough description of both CSC setups can be found in the "Technical Design Report for the Cryogenic Stopping Cell of the Super-FRS at FAIR" [1]. The Schematic of the prototype of the cryogenic stopping cell of the Super-FRS is shown in Figure 1.

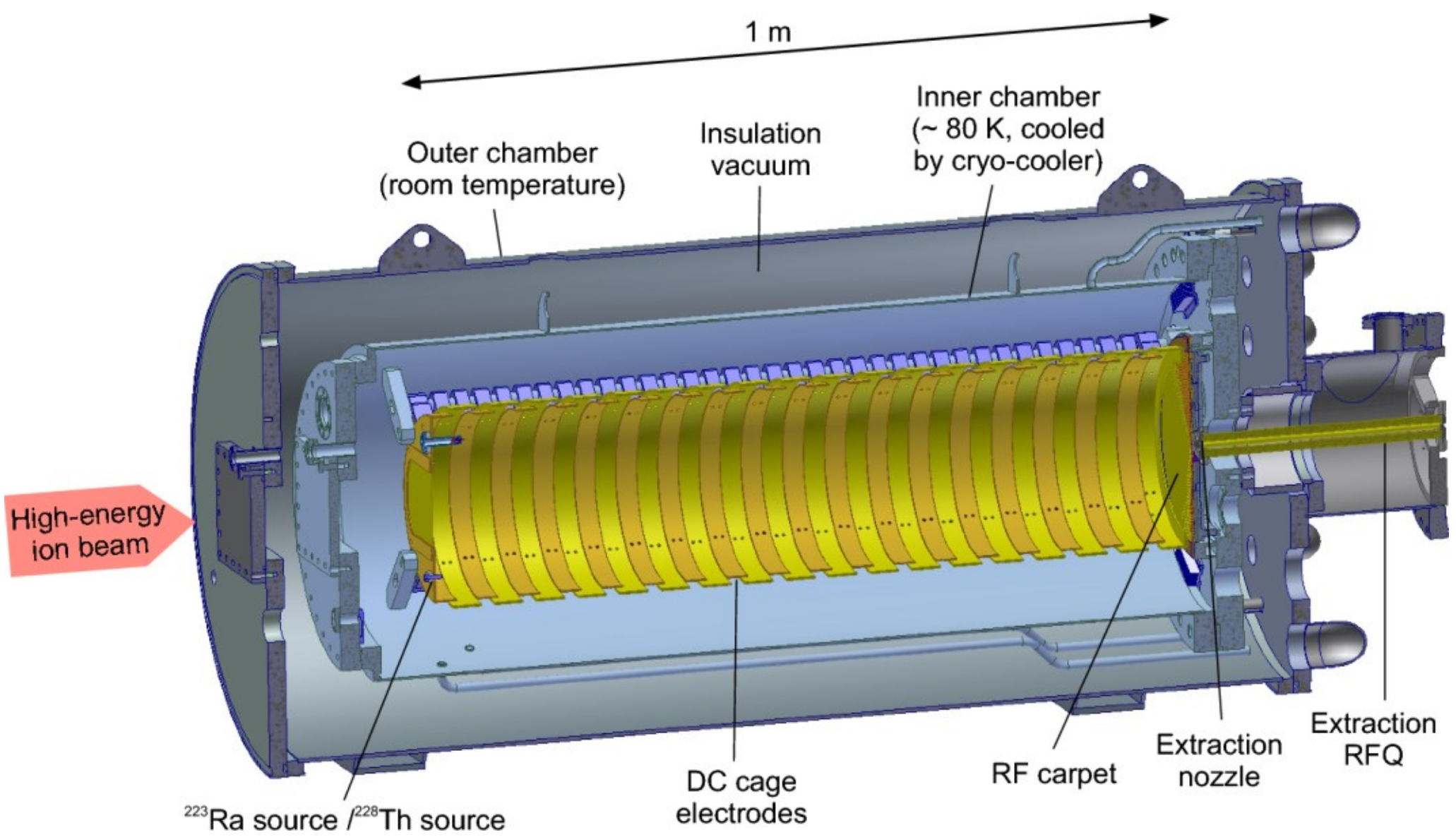

**Figure 1**. Schematic of the prototype of the cryogenic stopping cell of the Super-FRS. Reproduced from [1]. CC By 4.0.

The primary high-energy ion beam is injected into the cryogenic stopping cell via a thin metallic window, where it slows down and cools to the gas temperature through collisions with helium atoms. The ions are then transported through the cell to the radio frequency (RF) carpet via an applied DC electric field, as shown in Figure 1. In the vicinity of the RF carpet surface, the ions are focused towards its center by the combined action of the applied RF and DC fields. The slow, cooled continuous ion beam is extracted from the cell by a supersonic gas flow through a small-diameter extraction nozzle in the RF carpet. Finally, the ions are separated from the gas by passing through the extraction RF quadrupole (RFQ) under vacuum pumping conditions.

The conceptual design of the future cryogenic stopping cell of the Super-FRS is illustrated schematically in Figure 2. This CSC is composed of two chambers. The lower chamber, measuring 2 meters in length, is employed for the deceleration of the primary high-energy ion beam from the Super-FRS. As illustrated in Figure 2, the stopped ions are extracted perpendicularly to the direction of the primary ion beam as it passes through the nozzles in the separate RF carpets. The differentially pumped upper chamber allows low-energy ions to be extracted from the CSC. This is done through a cylindrical nozzle in the center of the large upper RF carpet. This extraction nozzle is of 0.6 mm in diameter and of 1.6 mm long. Then, the ions are carried by the gas jet into the extraction RFQ, which is placed on the axis near the nozzle exit plane.

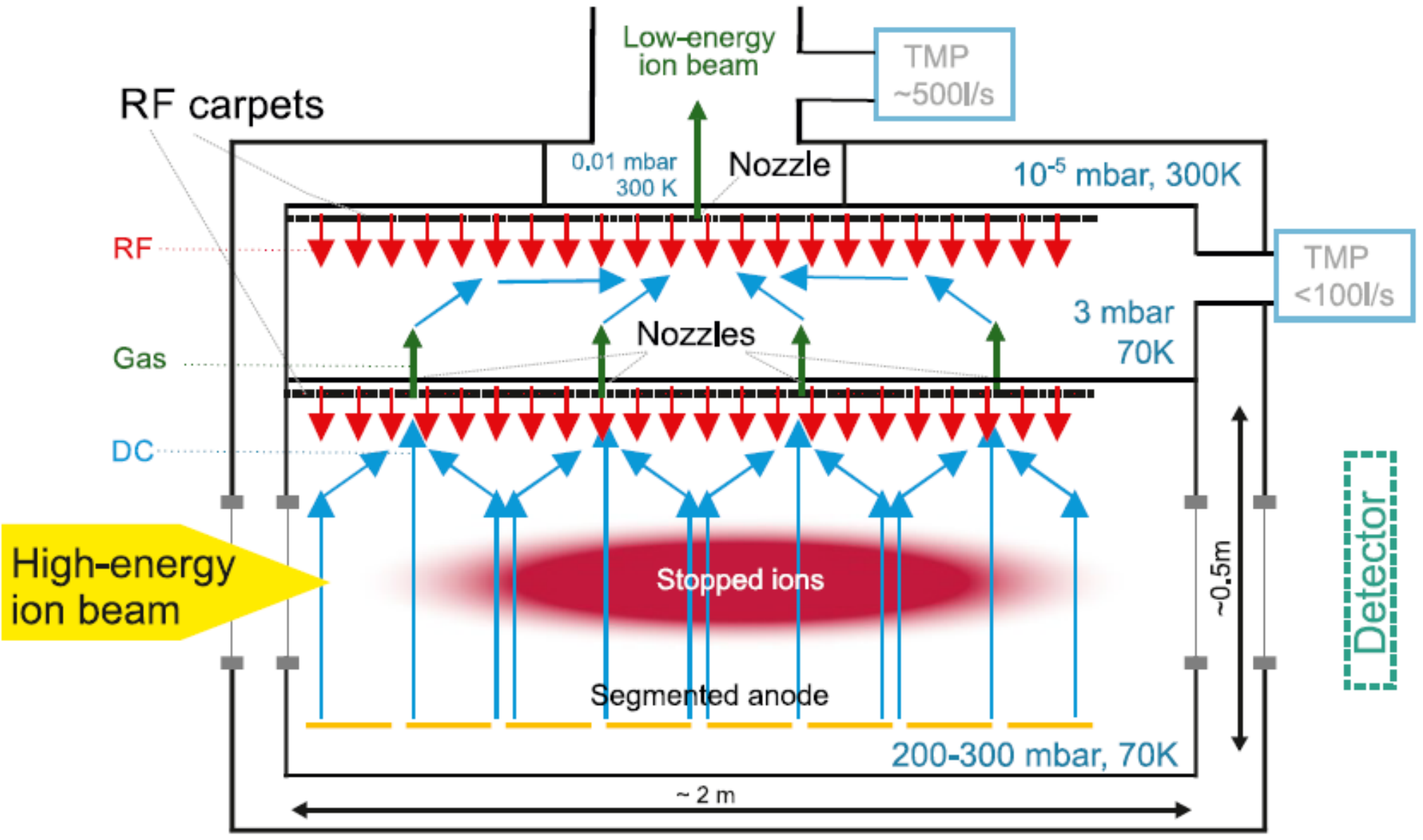

**Figure 2**. Schematic of the cryogenic stopping cell of the Super-FRS. Reproduced from [1]. CC By 4.0.

The RFQ technique is used a lot in many rare isotope facilities around the world. It's used for extracting ions from the gas cells, guiding ions, and converting continuous ion beams into short-pulsed ones. This article does not describe the RFQ technique, but readers can find this information in many other research papers (for example, see [1-22] and the references within).

In 2001, we proposed an original gas-dynamic technique for manipulating ion beams [23]. This technique facilitates the extraction of low-energy ion beams from gas stopping cells into a vacuum using RF-only electrodynamic funnels. It seems to be a highly competitive alternative to the current RFQ technique. Later, different versions of this gas-dynamic technique were created for various uses. You can find more information on this subject in other our publications, such as the review [24] and the original articles [25-27].

The ion beams that are continuously extracted by the gas flow from both CSCs do not need any additional cooling. However, they should be converted into pulesed beams before being delivered to the experimental setups. This issue can be easily solved by replacing the extraction RFQ (see Figure 1) with the short gas-dynamic cylindrical RF buncher, which has a similar design to the one described in [27].

For example, the gas-dynamic RF buncher device was used to create the original laser ablation ion source at TU Darmstadt [28, 29]. For the illustration, check out the technical drawing of this ion source with an ablation unit, along with photos of the RF-only funnel and RF buncher shown in Figure 7 in our previous paper [27], which we reproduced from [30].

We have investigated the proposed here gas-dynamic RF buncher using detailed gas-dynamic and ion-trajectory Monte Carlo simulations. The results of these simulations are presented and discussed in the next sections.

The earlier similar simulations and their respective outcomes have been documented in other sources (see references [24–30]).

Detailed gas dynamic simulations of the buffer gas flow we have made using the VARJET code. This code is based on the solution of a full system of time dependent Navier–Stokes equations and is described in detail in [31]. The results of the gas-dynamic simulations (flow fields of the buffer gas velocity, density and temperature) were then used in a Monte Carlo ion-trajectory simulations, where the electric fields were simulated using SIMION 8.1 [32]. Note that our computational results (gas dynamic + Monte Carlo) for ions extracted from the gas jet into vacuum are in good agreement with the measurements. Based on our gas-dynamical calculations, the installations of gas-jet internal targets [33, 34] have been created. The results of these calculations are also in good agreement with the measured parameters and structure of the supersonic gas jet.

## 2. The future cryogenic stopping cell of the Super-FRS

### 2.1 Gas dynamic simulations

The gas dynamic and ion trajectory simulations were performed for the following set of design and boundary parameters, which have been used for simulations in [1] (Section "5.3 Ion transport from stopping region to extraction region"):

- cylindrical nozzle in the center of the large upper RF carpet (see Figure 2) is of 0.6 mm in diameter and of 1.6 mm long;
- helium pressure and temperature in the extraction region are 3 mbar and 75 K, correspondingly;
- effective pumping speed of turbomolecular pump is about of 400 l/s (the nominal pumping speed is of 500 l/s.

The cylindrical RF buncher is composed of 10 thin metal electrodes and has a total length of 6 mm. The electrodes have aperture of 2 mm, a thickness of 0.1 mm, and a spacing of 0.5 mm between adjacent electrodes.

Figure 3 shows the design of the RF buncher, along with the results of a detailed gas dynamic simulation for the helium density flow field.

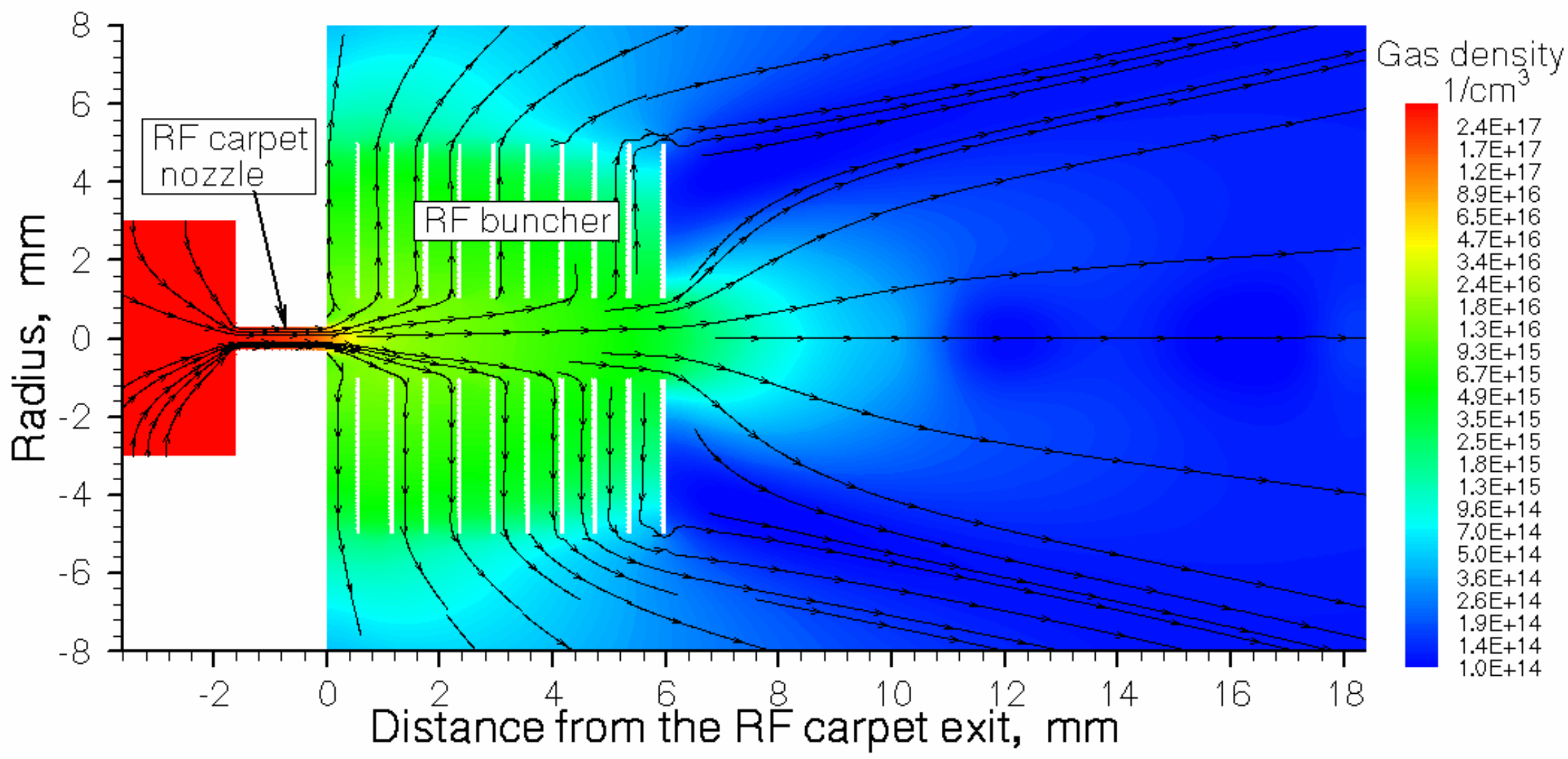

**Figure 3.** Schematic view of the RF buncher combined with the results of the gas-dynamic simulation for the helium density flow field. The stagnation input gas pressure and temperature are 3 mbar and 75 K, respectively. Black arrowed lines indicate the direction of gas flow.

The Figure 4 and Figure 5 illustrates the data of Figure 3 for the helium velocity and density distributions along the axis, correspondingly.

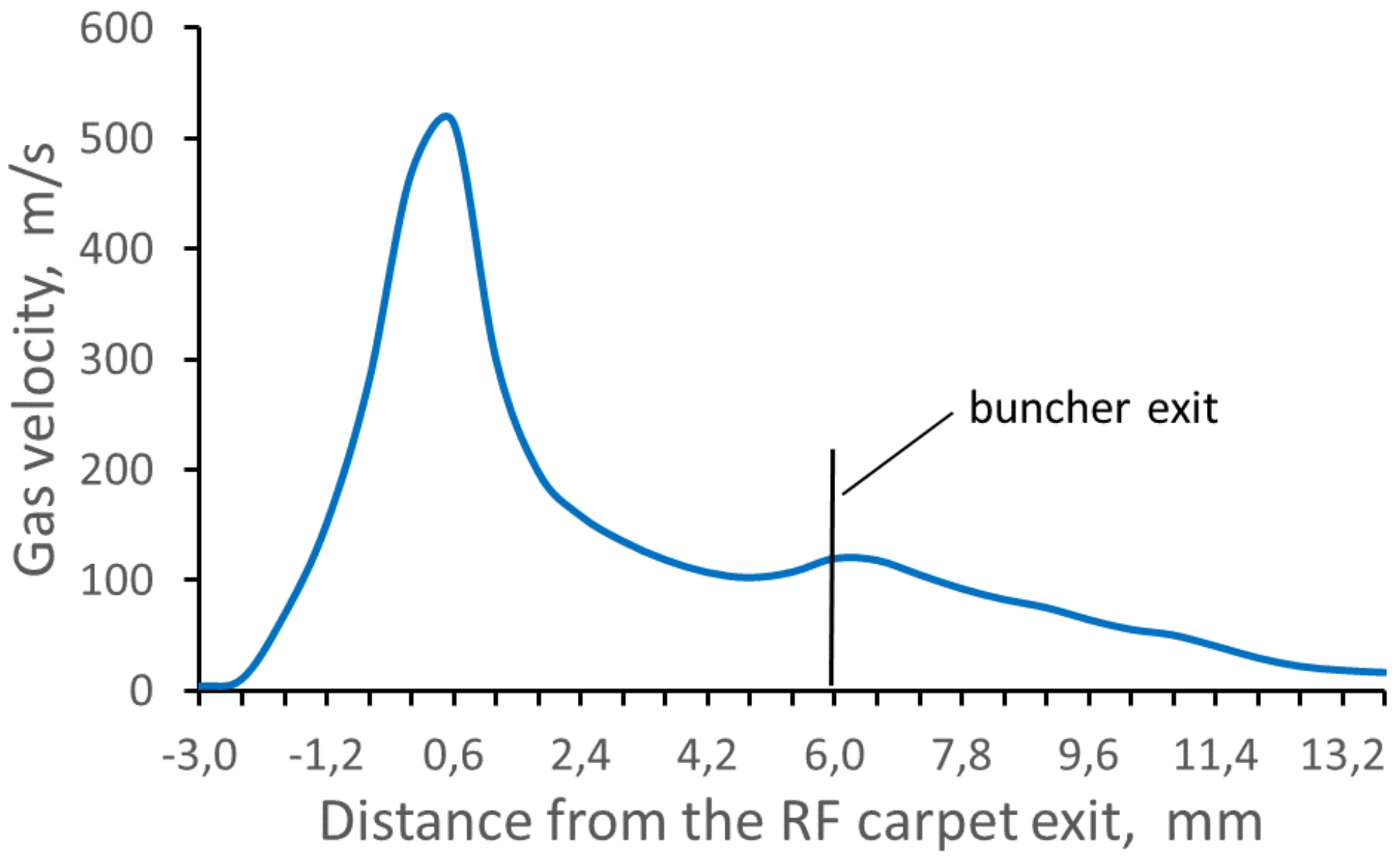

**Figure 4.** The calculated helium velocity distribution along the axis. It illustrate the results of gas-dynamic simulation shown in Figure 3, as well.

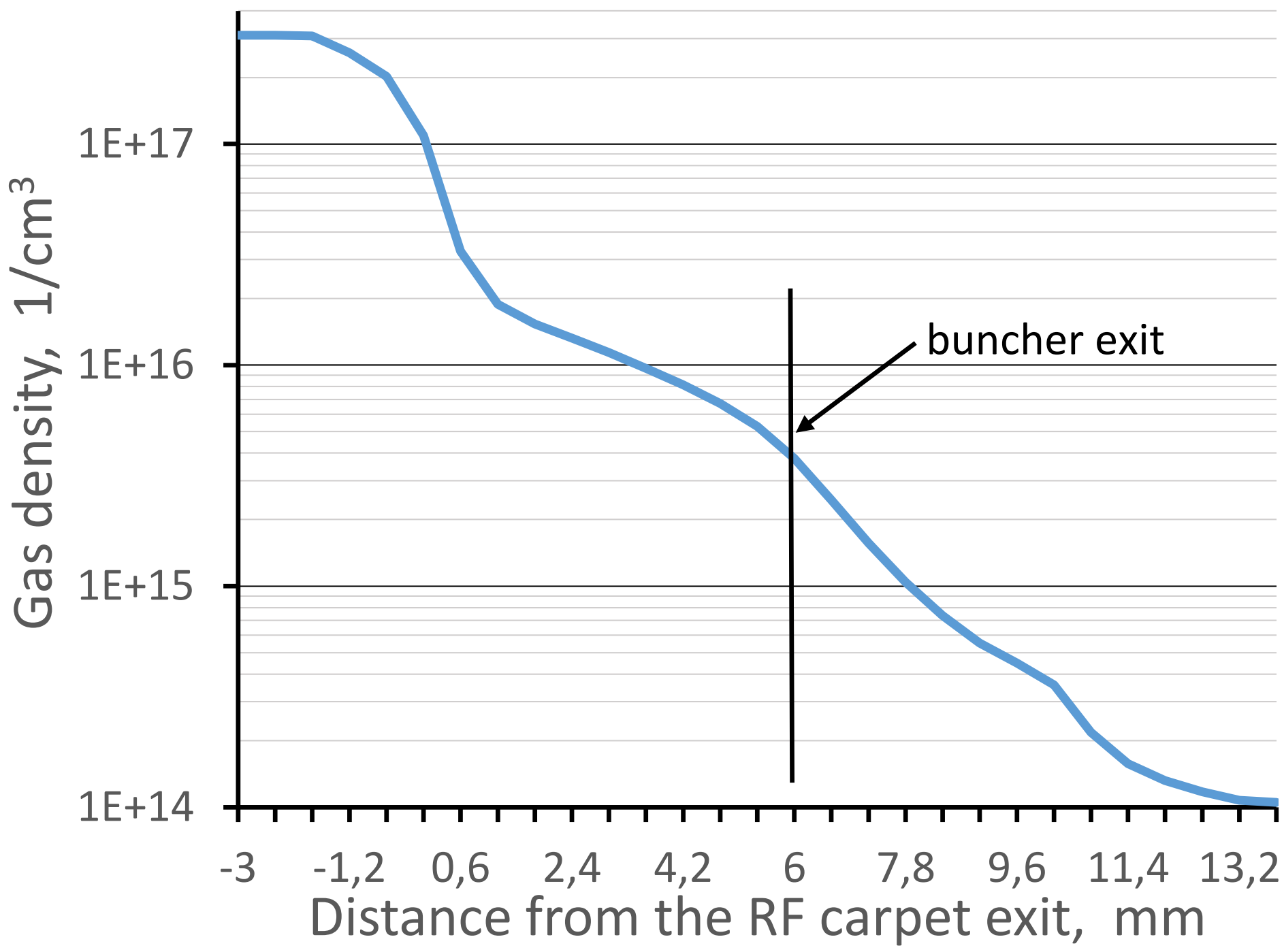

**Figure 5.** The calculated helium density distribution along the axis. It illustrate the results of gas-dynamic simulation shown in Figure 3, as well.

The calculated helium flow rate through the RF carpet nozzle is of 0.35 mbar l/s (for the 300 K temperature of vacuum pump).

### 2.2 Ion trajectory simulations

Monte Carlo ion trajectory simulations were performed for both continuous and pulsed ion beam modes. The trajectory of each ion is calculated from the exit of the RF carpet nozzle to a distance of 10 mm downstream from the exit of the gas-dynamic RF buncher, as illustrated in Figure 3.

In order to create a potential well for the ion bunching inside the RF buncher, a positive electric potential is applied to the final buncher electrode. Note that it is not necessary to apply any RF voltage to this trapping electrode. We found that, for optimal ion bunching, the trapping voltage should be proportional to the direct current (DC) electric field strength that is applied to the buncher electrodes. For example, at a DC electric field strength of 2 V/mm, the trapping potential should be about +50 V; at an electric field strength of 1 V/mm, it should be +25 V, and so on. With a DC electric field gradient of 1.0 V/mm, the voltage difference between two adjacent buncher electrodes is 0.6 V. It is important to note that all Monte Carlo simulations were conducted at a constant RF frequency of 10 MHz. The RF voltage is applied to the buncher electrodes in such a manner that the phases between adjacent electrodes are opposite.

Under these conditions, ions of different masses are squeezed into a small-sized bunch at the position of the seventh buncher electrode (at 4.2 mm from the buncher entrance), as illustrated in Figure 3. The ion bunch is shaped like a cylinder, with a diameter of about 0.8 mm and a length of 0.1 mm. The the thickness of the electrodes is also 0.1 mm. The gas density at this location inside the RF buncher is $8.1\times10^{15}$ atoms/cm³ (see Figure 5).

The results of the Monte Carlo simulations for the longitudinal velocity distribution of the extracted ion beam with a mass of 100 u at various trapping times are shown in Figure 6.

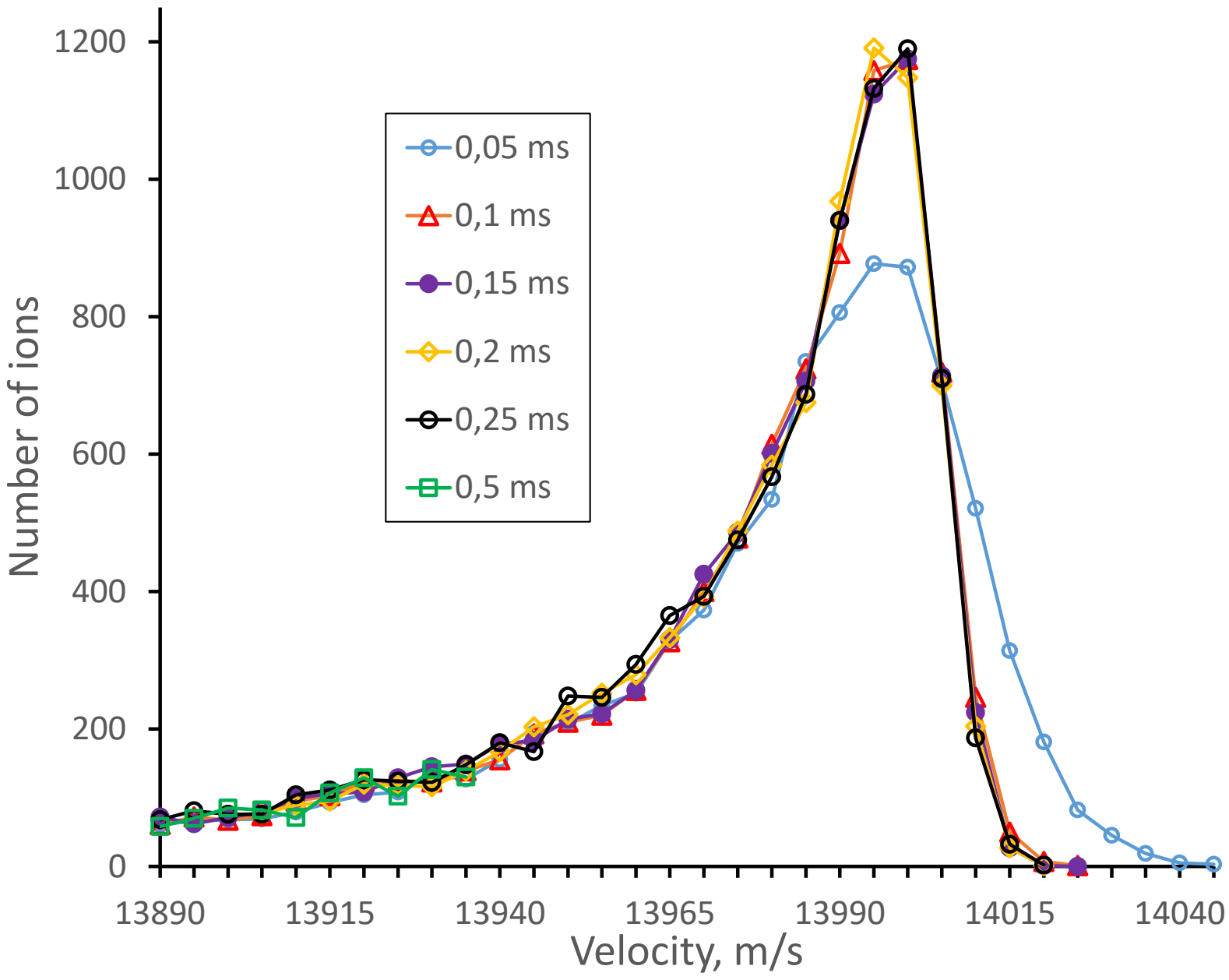

**Figure 6**. The results of the Monte Carlo ion-trajectory simulations for the temporal distribution of the extracted ion pulse for different trapping times. The values on the time axis correspond to the time-of-flight of minus the trapping time. The ion mass is 100 u. The DC electric field strength are 1.0 V/mm inside the buncher and 10 V/mm at extraction region behind the buncher exit. The RF voltage is 125 $V_{pp}$ (peak-to-peak). For each case of the trapping time, the number of extracted ions is 10 000.

Figure 7 shows the results of the Monte Carlo simulations of the temporal distribution of the extracted 100 u ion beam at various trapping times. The values on the time axis correspond to time-of-flight minus trapping time. This provides a better understanding of the ion bunching process.

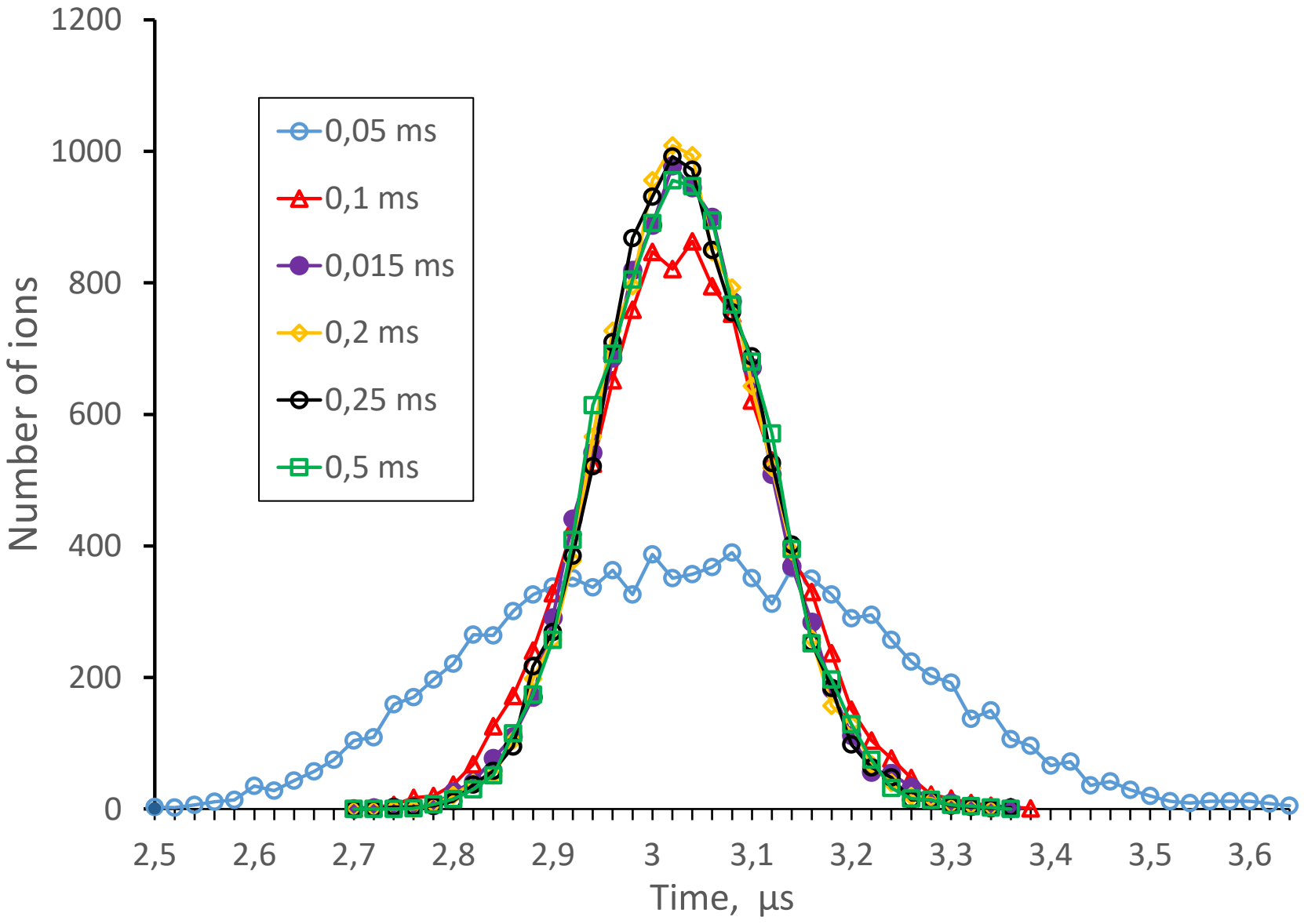

**Figure 7**. The results of the Monte Carlo ion-trajectory simulations show the longitudinal velocity distribution of the extracted ion pulse for different trapping times. The values on the time axis correspond to the time-of-flight minus the trapping time. The ion mass is 100 u. The DC electric field strength are 1.0 V/mm inside the buncher and 10 V/mm at extraction region behind the buncher exit. The RF voltage is 125 $V_{pp}$. For each case of the trapping time, the number of extracted ions is 10,000.

As Figures 6 and 7 show, a trapping time of 0.1 ms is sufficient for ions to reach thermal equilibrium with the gas inside the gas-dynamic RF buncher. The distance between the 10th trapping electrode and the 7th electrode (where the bunch is located) is 1.8 mm. This means that about 2 µs is enough time to extract all bunched ions from the trapping well into a vacuum. Because of this, the RF buncher can effectively operate at a repetition pulse rate higher than 1 kHz. This is possible by periodically quickly switching off the positive voltage applied to the exit RF bunches electrode for 2 µs.

The results of the Monte Carlo simulations for the characteristics of extracted from gas-dynamic RF buncher pulsed ion beams of different masses are shown in Table 1. It is important to note here that all ions with a mass greater than M = 20 u are transmitted through the RF buncher with 100% efficiency. The transmission efficiency of the ions with M = 20 u is about 95%.

**Table 1.** Calculated characteristics of the extracted into a vacuum ion beams of different masses are shown for the **pulsed-beam** mode. The DC electric field strengths are 1.0 V/mm inside the buncher and 10 V/mm in the extraction region behind the buncher exit. The applied RF frequency is 10 MHz. The acronym FWHM signifies Full-Width at Half-Maximum.

| **Extracted ion beam characteristic** | **Ion mass** | | | | |
|---|---|---|---|---|---|
| | **20** | **40** | **100** | **150** | **200** |
| Longitudinal velocity (m/c) | 31310 | 22130 | 14000 | 11430 | 9895 |
| Longitudinal (FWHM) velocity spread (m/c) | 87.4 | 44.6 | 24.9 | 18.5 | 16.5 |
| Transverse velocity (m/s) | 316 | 250 | 163 | 148 | 127 |
| Transverse (FWHM) velocity spread (m/s) | 431 | 323 | 159 | 130 | 113 |
| Beam radius (90%) (mm) | 0.69 | 0.65 | 0.54 | 0.54 | 0.53 |

| Transverse emittance $\varepsilon_{x,y}$ ($\pi$·mm·mrad) | 8.50 | 7,34 | 6.9 | 7,0 | 7,0 |
|---|---|---|---|---|---|
| Normalized emittance $\varepsilon^{N}_{x,y} = \varepsilon_{x,y} \cdot [E]^{1/2}$ ($\pi$·mm·mrad ·$[eV]^{1/2}$) | 84.0 | 73,4 | 68.8 | 68,9 | 68,9 |
| Bunch time (FWHM) width (µs) | 0,15 | 0,21 | 0,18 | 0,22 | 0,23 |
| Longitudinal emittance (eV·ns) | 0,30 | 0,23 | 0,12 | 0,12 | 0,13 |
| Applied RF voltage (peak-to-peak) | 70 | 80 | 125 | 150 | 175 |

## 3. The prototype of the cryogenic stopping cell in the FRS at GSI

The ions are flowing out of the current CSC at the FRS via a supersonic gas jet into the "extraction RFQ" as shown schematically in Figure 1 and shortly described above in the Introducnion. The operation of this CSC at helium stagnation pressure of 100 mbar at a temperature of 80 K (see Section 3.2.2 in [1]) results in a background gas pressure in the extraction region of approximately $1\times10^{-2}$ mbar. This vacuum level is supported by a set of three turbomolecular pumps, with a total pumping speed of up to 4000 l/s. The gas flow rate through the cylindrical nozzle, with a diameter of 0.6 mm and length of of 1.6 mm at the center of the RF carpet, is equivalent to 23.1 mbar l/s at a 300 K temperature of the vacuum pumps.

This calculated gas flow rate through the nozzle is 66 times higher than that of the future CSC of the Super-FRS. The gas-dynamic RF buncher described above, which has 10 thin electrodes, clearly could not reach the goal of properly extracting and bunching ion beams. However, we discovered that increasing the number of electrodes to 50, while maintaining their original design, would also solve this problem. It is important to emphasize that you do not need to apply the DC field strength to the first 40 electrodes because the gas-jet transports the ions quickly and effectively. In other words, this part of the gas-dynamic RF buncher works like the RF-only funnels described in detail in our review [24].

### 3.1 Gas dynamic simulations

Figure 8 shows the design of the RF buncher that consists of 50 thin electrodes, along with the results of a detailed gas dynamic simulation for the helium velocity flow field.

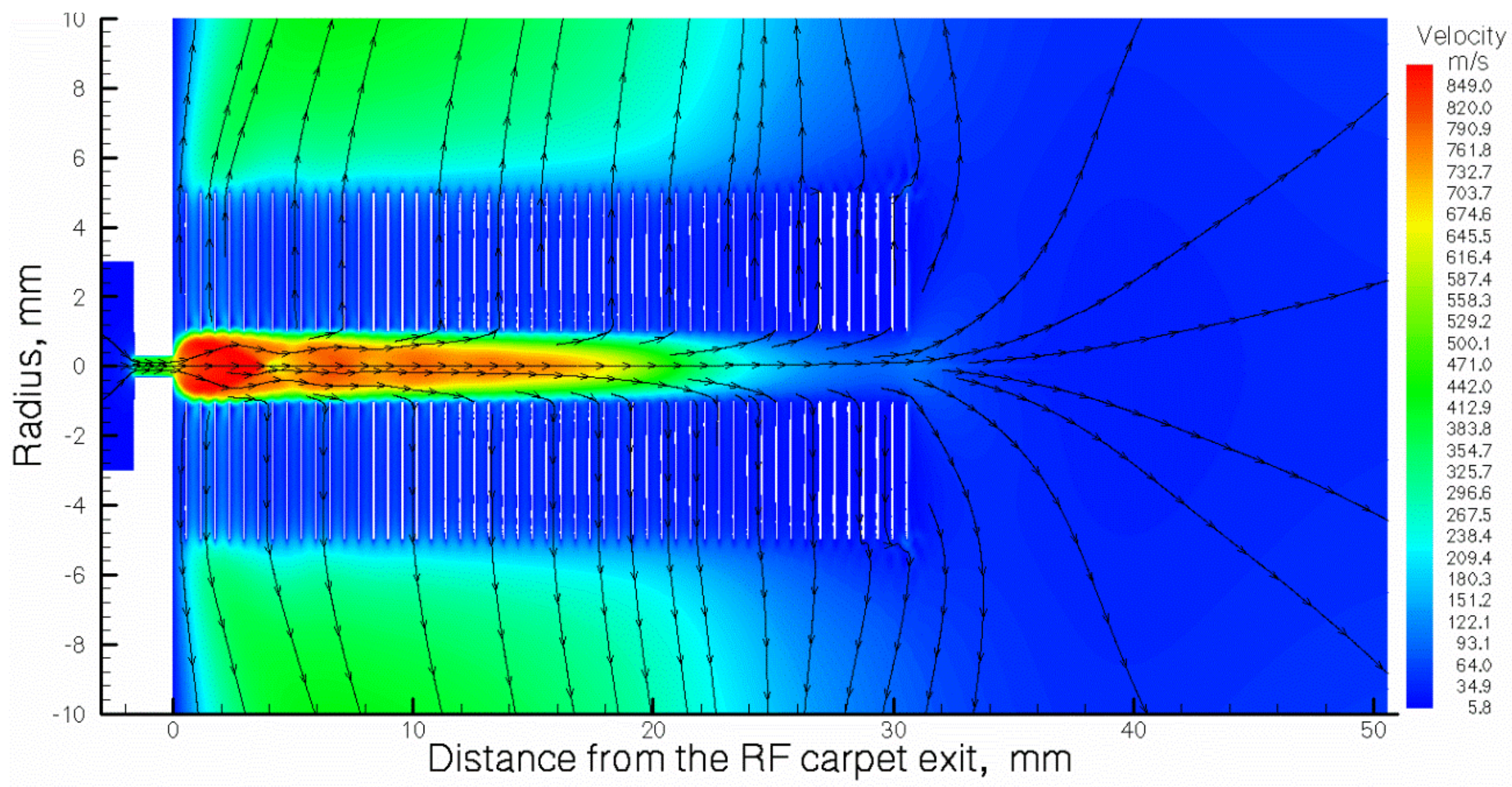

**Figure 8.** Schematic view of the RF buncher that consists of 50 thin electrodes combined with the results of the gas-dynamic simulation for the helium velocity flow field. The stagnation input gas pressure and temperature are 100 mbar and 80 K, respectively. Black arrowed lines indicate the direction of gas flow.

The shock wave barrel structures of the supersonic jet inside the RF buncher, along with an intense gas flow through the gaps between the electrodes into a vacuum, are clearly visible in Figure 8.

The Figure 9 and Figure 10 illustrates the data of Figure 8 for the helium velocity and density distributions along the axis, correspondingly.

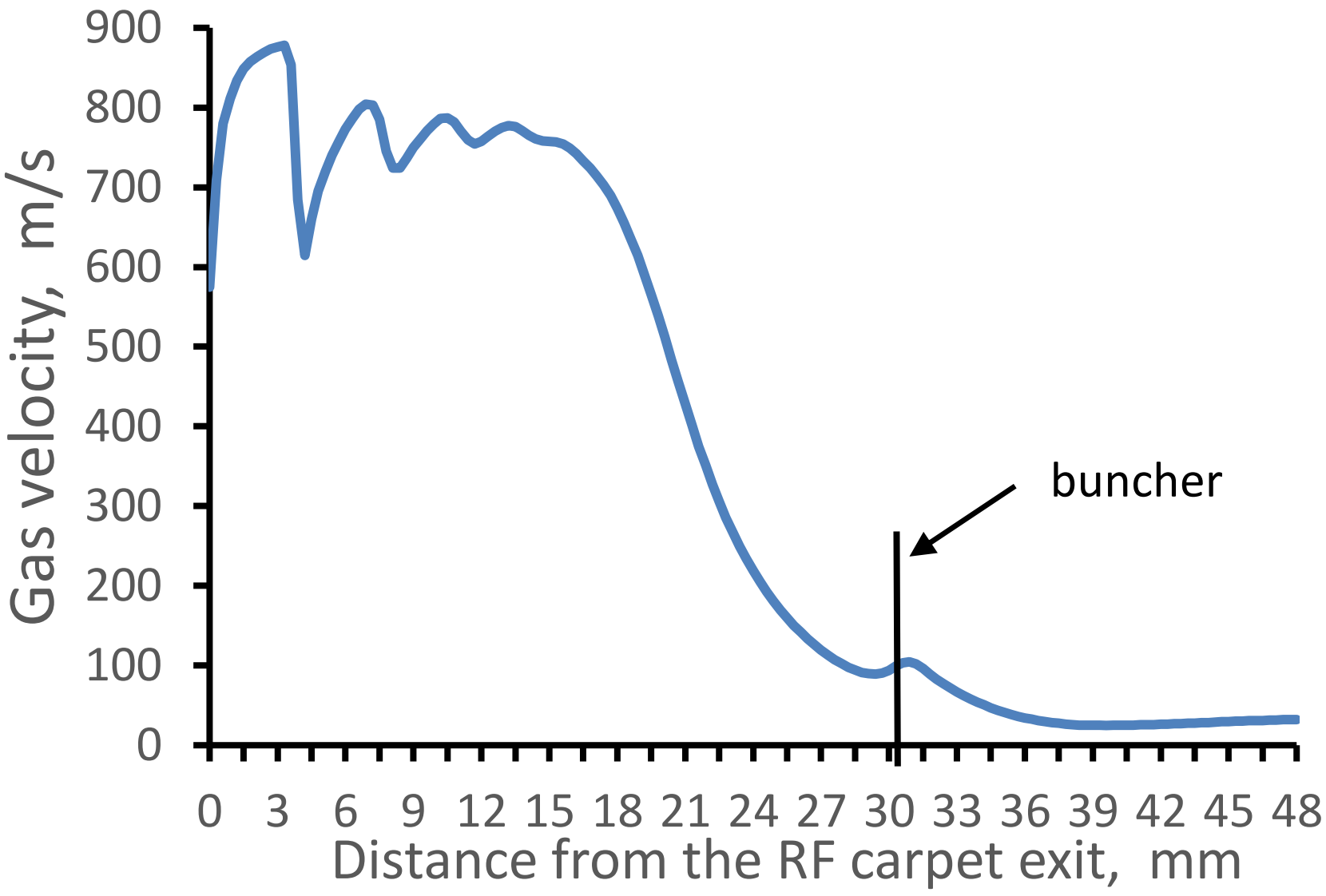

**Figure 9.** The calculated helium velocity distribution along the axis. It illustrate the results of gas-dynamic simulation shown in Figure 8, as well.

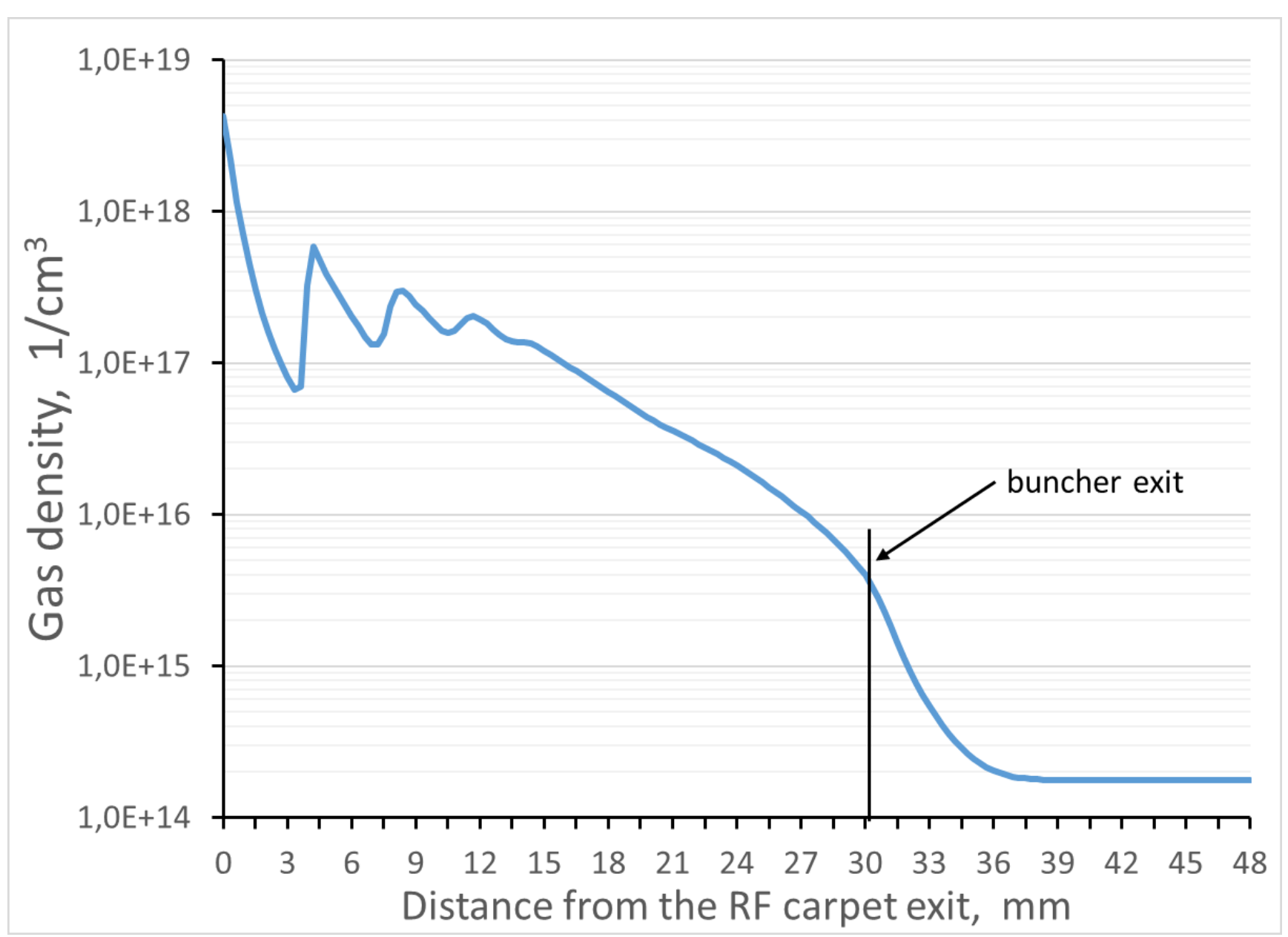

**Figure 10.** The calculated helium density distribution along the axis. It illustrate the results of gas-dynamic simulation shown in Figure 8, as well.

### 3.2 Ion trajectory simulations

The results of the Monte Carlo simulations for the characteristics of the extracted pulsed ion beams with a mass of 100 u from the long gas-dynamic RF buncher (that we propose for the prototype of the CSC installed in the FRS at GSI) are shown in Table 2. For comparison, similar data are also shown here for the short RF buncher for the future CSC in the Super-FRS. In both RF buncher variants, the small-sized ion bunches exhibit the same shape and position at the third electrode from the buncher's exit.

**Table 2**. The calculated characteristics of the pulsed ion beams extracted from the **future CSC** in the Super-FRS and the **current CSC** in the RFS. The ion mass is 100 u. The DC electric field strength are 1.0 V/mm inside the buncher and 10 V/mm at extraction region behind the buncher exit. The RF voltage and RF frequency are 125 $V_{pp}$ and 10 MHZ, correspondingly. The trapping time is 0.25 ms and the number of extracted ions is 10,000.

| Calculation variant | **future CSC** | **current CSC** |
|---|---|---|
| Longitudinal velocity (m/c) | 14000 | 13910 |
| Longitudinal (FWHM) velocity spread (m/c) | 24.9 | 18.9 |
| Transverse velocity (m/s) | 134 | 134 |
| Transverse (FWHM) velocity spread (m/s) | 179 | 176 |
| Beam radius (90%) (mm) | 0.54 | 0.535 |
| Transverse emittance $\varepsilon_{x,y}$ ($\pi$·mm·mrad) | 6.85 | 6.47 |
| Normalized emittance $\varepsilon^N_{x,y} = \varepsilon_{x,y}\cdot[E\ ]^{1/2}$ ($\pi$·mm·mrad ·[eV]$^{1/2}$) | 67.7 | 63.2 |
| Bunch time (FWHM) width (μs) | 0.18 | 0.18 |
| Longitudinal emittance (eV·ns) | 0.12 | 0.093 |

Figure 11 illustrate the data in Table 2 for the temporal distribution of the extracted ion pulse for the both CSCs.

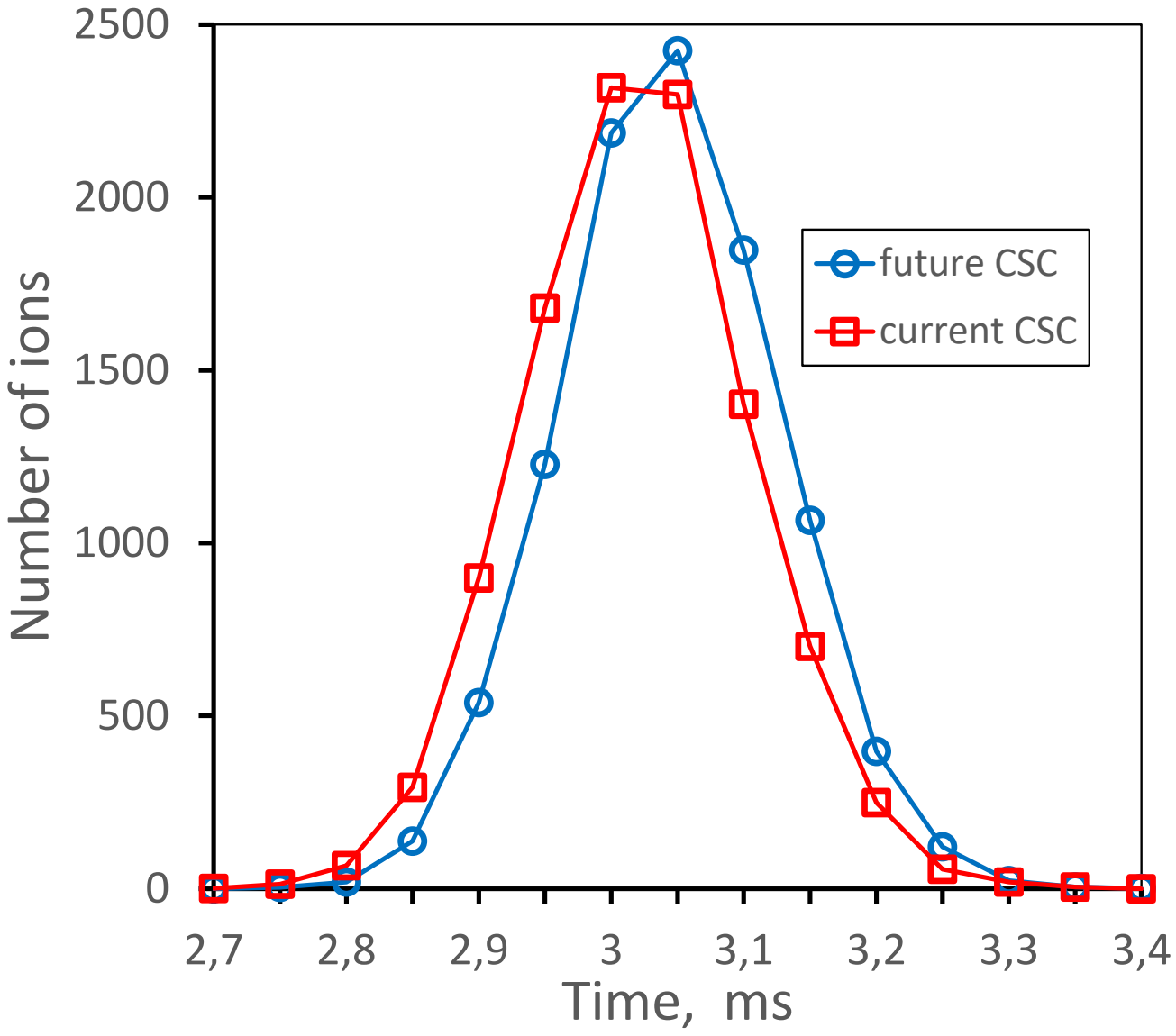

**Figure 11**. The results of the Monte Carlo simulations for the temporal distribution of the pulsed ion beams extracted from the both CSCs. The mass of an ion beam is 100 u. The values on the time axis correspond to the time-of-flight minus the trapping time. The DC electric field strengths are 1.0 V/mm inside the buncher and 10 V/mm in the extraction region behind the buncher exit. The RF voltage and RF frequency are 125 $V_{pp}$ and 10 MHZ, correspondingly. The trapping time is 0.25 ms and the number of extracted ions is 10,000.

As shown in Table 2 and Figure 11, the characteristics of the pulsed ion beams are very similar for the two CSCs. This is explained by the similarity of the gas density flow fields in the region of the last ten electrodes of the both RF bunchers (see Figures 5 and 10). For example, the gas densities at the bunch location inside the RF buncher are $8.1 \times 10^{15}$ atoms/cm$^3$ and $6.9 \times 10^{15}$ atoms/cm$^3$ for the future and the current CSC, respectively.

## 4. The extraction of continuous ion beams

The results of the Monte Carlo simulations for the extraction of continuous ion beams from both FAIR CSCs with the use of the gas-dynamic RF bunchers described above are presented in Table 3. In order to organise the continuous beam operation mode, it is necessary only to switch off the trapping voltage at the last RF buncher electrode.

**Table 3**. The calculated characteristics of the continuous ion beams extracted from the **future CSC** in the Super-FRS and the **current CSC** in the RFS. The mass of an ion beam is 100 u. The DC electric field strengths are 1.0 V/mm inside the buncher and 10 V/mm in the extraction region behind the buncher exit. The RF voltage and RF frequency are 125 $V_{pp}$ and 10 MHZ, correspondingly. The number of extracted ions for each calculation variant is 20,000.

| Calculation variant | **future CSC** | **current CSC** |
|---|---|---|
| Longitudinal velocity (m/c) | 14170 | 14110 |
| Longitudinal (FWHM) velocity spread (m/c) | 116 | 117.6 |
| Transverse velocity (m/s) | 134 | 134 |
| Transverse (FWHM) velocity spread (m/s) | 217 | 205 |
| Beam radius (90%) (mm) | 0.36 | 0.35 |
| Transverse emittance $\varepsilon_{x,y}$ ($\pi$·mm·mrad) | 3.41 | 3.33 |
| Normalized emittance $\varepsilon^N_{x,y} = \varepsilon_{x,y} \cdot [E]^{1/2}$ | 34.1 | 33.20 |
| ($\pi$·mm·mrad ·[eV]$^{1/2}$) | | |
| Total time of flight (FWHM) [μs] | 4.7 | 38.7 |

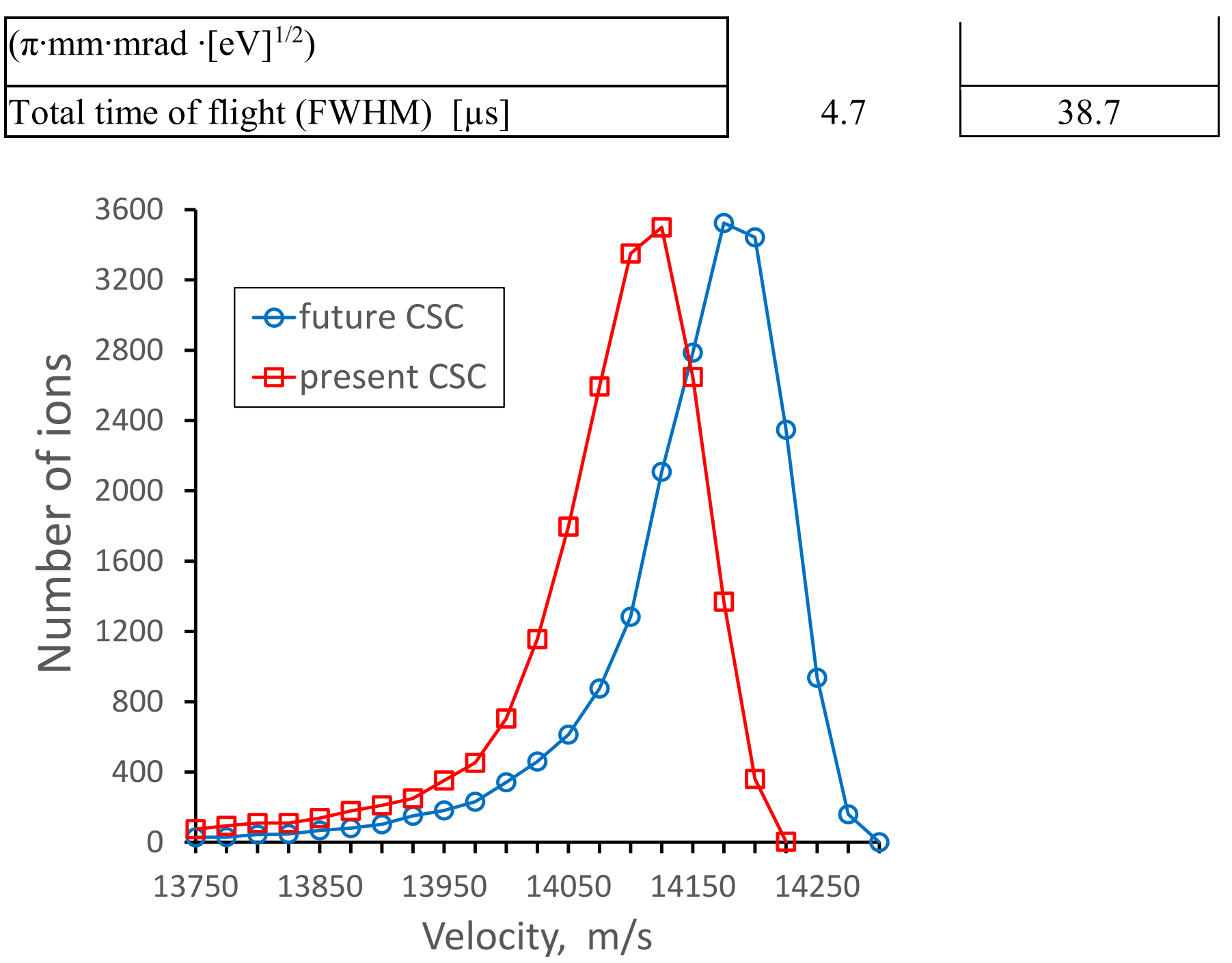

**Figure 12**. The results of the Monte Carlo simulations for the longitudinal velocity distribution of the continuous ion beams extracted from the **future CSC** of the Super-FRS and the **current CSC** of the RFS. The mass of an ion beam is 100 u. The DC electric field strengths are 1.0 V/mm inside the buncher and 10 V/mm in the extraction region behind the buncher exit. The RF voltage and RF frequency are 125 $V_{pp}$ and 10 MHZ, correspondingly. The number of extracted ions for each calculation variant is 20,000.

## 5. A brief comparison of the RFQ and gas-dynamic technique for ion beam cooling and bunching.

Let us compare the performance of the gas-dynamic technique described above for ion-beam extraction and bunching in cryogenic stopping cells at FAIR with that of a typical RFQ technique. As mentioned in Section 2.2, helium density at the location of the trapped ions inside the RF buncher of the future CSC is $8.1\times10^{15}$ atoms/cm$^3$. It corresponds to the mean free path length of about 0.6 mm. The value of the thermal velocity of the trapped ions with a mass of 100 u at a helium temperature of 75 K is about 112 m/s. This means that the frequency of collisions between the ions and helium atoms is $1.87\times10^5$ Hz. During a 0.25 ms trapping time, the ions undergo an average of 47 collisions. This is sufficient for the ions to become thermalized in the gas.

In the case of traditional RFQ cooler bunches, which usually operate at a helium pressure of about $1\times10^{-2}$ mbar and at room temperature, the mean free path length is 20.4 mm. This corresponds to an average collision frequency of $1.1\times10^4$ Hz for ions with a mass of 100 u. So, on average, these ions only collide 2.3 times during the same 0.25 ms. This is clearly not enough to reach a state of thermal equilibrium with gas.

It explains why the existing RFQ cooler-bunchers require much longer trapping times for proper ion bunching. Additionally, the RFQ cooler-bunchers should be considerably longer, typically between 0.5 and 1 meters, to maintain sufficient gas density for effective ion cooling and bunching while preventing deterioration of the high vacuum in the external beamlines.

## 6. Discussion and outlook

This paper presents a proposal for the application of a novel gas-dynamic ion-beam extraction and bunching technique to enhance the quality of pulsed cold ion beams extracted from two cryogenic stopping cells at the FAIR.

Through detailed gas-dynamic and ion trajectory computer simulations, we demonstrated that implementing this proposal would enable achieving world-record emittance values for ion beams across a broad mass range. For example, the predicted longitudinal emittance value of 0.12 eV·ns in our simulation looks fantastic compared to the expected value of 500 eV·ns for the future CSC in the Super-FRS at FAIR (see Table 2 in [1]).

However, it is true and feasible because the solution proposed here is more than just a significant technical improvement to the RFQ method. Rather, it is an example of a qualitatively new technology in the field of low-energy ion beam manipulation.

The realization of the proposed ion beam extraction-bunching devices for both CSCs at FAIR can be achieved quickly and with minimal investments because they have simple designs and do not require additional equipment, such as vacuum chambers, pumps and so on.

There is no need to wait when the future CSC in the Super-FRS to be constructed and made available for use. The necessary experimental tests can be done offline at the existing CSC in the FRS using the radioactive ion source that is already there. You can see this ion source schematically illustrated in Figure 1.

There are many reasons to believe that the novel gas-dynamic technique presented here will allow for the replacement of the traditional RFQ cooler-bunchers currently in use, as well as those still under development, in laboratories where the gas stopping cells are used. For example, the gas-dynamic RF buncher described in Section 2 and shown in Figure 3 can be placed behind the RF carpet of the St. Benedict setup at Notre Dame University [19].

**Funding:** This research received no external funding.
**Data Availability Statement:** The data presented in this study are available upon request from the author.
**Conflicts of Interest:** The author declares no conflicts of interest

## References

[1] Daler Amanbayev, Wolfgang R. Plaß, Timo Dickel, Samuel Ayet San Andr´es, Peter Dendooven15, Hans Geissel, Emma Haettner, Muhsin N. Harakeh, Ivan Miskun, Sivaji Purushothaman, Manisha Ranjan, Moritz P. Reiter, Ann-Kathrin Rink, Christoph Scheidenberger, Zoran Andelkovic, Soumya Bagchi, Dimiter Balabanski, S¨onke Beck, Julian Bergmann1, Michael Block, Paul Constantin, Sterian Danaila, Jens Ebert, Florian Greiner, Lizzy Gr¨of, Fabian Heiße, Frank Herfurth, Christine Hornung, Christian Jesch, Nasser Kalantar-Nayestanaki, Israel Mardor, Mihai Merisanu, Iain D. Moore, Martin Petrick, Stephane Pietri, Ilkka Pohjalainen, Anamaria Spataru, Alexandru N. State, Helmut Weick, John S. Winfield, Martin Winkler, Mikhail I. Yavor, and for the Super-FRS Experiment Collaboration, Technical design report for the cryogenic stopping cell of the Super-FRS at FAIR, Eur. Phys. J. Spec. Top. (2025) 234:861–917, https://doi.org/10.1140/epjs/s11734-025-01605-z.

[2] A. Jaries , J. Ruotsalainen, R. Kronholm, T. Eronen, A. Kankainen, HIBISCUS: a new ion beam radio-frequency quadrupole cooler-buncher for high-precision experiments with exotic radioactive ions, Nuclear Instruments and Methods in Physics Research Section A 1073 (2025) 170273, https://doi.org/10.1016/j.nima.2025.170273.

[3] A. Nieminen, J. Huikari, A. Jokinen, J. Äystö, P. Campbell, E. Cochrane, Beam cooler for low-energy radioactive ions, Nuclear Instruments and Methods in Physics Research Section A 469 (2001) 244–253, http://dx.doi.org/10.1016/S0168-9002(00)00750-6.
[4] A. Nieminen, P. Campbell, J. Billowes, D. H. Forest, J. A. R. Griffith, J. Huikari, A. Jokinen, I. D. Moore, R. Moore, G. Tungate, J. Äystö, On-Line Ion Cooling and Bunching for Collinear Laser Spectroscopy, Phys. Rev. Lett. 88, 094801, Feb 2002.
[5] I.D. Moore, P. Dendooven, J. Ärje, The IGISOL technique—three decades of developments, Hyperfine Interact. 223 (1) (2014) 17–62, http://dx.doi.org/10.1007/s10751-013-0871-0 .
[6] I.D. Moore, New concepts for the ion guide technique, Nuclear Instruments and Methods in Physics Research Section B, 266 (2008) 4434.
[7] F. Herfurth, J. Dilling, A. Kellerbauer, G. Bollen, S. Henry, H.-J. Kluge, E. Lamour, D. Lunney, R.B. Moore, C. Scheidenberger, S. Schwarz, G. Sikler, J. Szerypo, A linear radiofrequency ion trap for accumulation, bunching, and emittance improvement of radioactive ion beams, Nuclear Instruments and Methods in Physics Research Section A 469 (2001) 254–275.
[8] G. Savard, A.F. Levand, B.J. Zabransky, The CARIBU gas catcher, Nuclear Instruments and Methods in Physics Research Section A 685 (2012) 70, https://doi.org/10.1016/j.nima.2012.05.031.
[9] Ramzi Boussaid, G. Ban, G. Quéméner, Y. Merrer, and J. Lorry, Development of a radio-frequency quadrupole cooler for high beam currents, Phys. Rev. Accel. Beams, 20, (2017) 124701, https://doi.org/10.1103/PhysRevAccelBeams.20.124701.
[10] B.R. Barquest, G. Bollen, P.F. Mantica, K. Minamisono, R. Ringle, S. Schwarz, RFQ beam cooler and buncher for collinear laser spectroscopy of rare isotopes, Nuclear Instruments and Methods in Physics Research Section A, 866 (2017) 18, https://doi.org/10.1016/j.nima.2017.05.036.
[11] S. Schwarz, G. Bollen, R. Ringle, J. Savory, P. Schury, The LEBIT ion cooler and buncher, Nuclear Instruments and Methods in Physics Research Section A, 816 (2017) 131, https://doi.org/10.1016/j.nima.2016.01.078.
[12] H. Franberg, P. Delahaye, J. Billowes, K. Blaum, R. Catherall, F. Duval, O. Gianfrancesco, T. Giles, A. Jokinen, M. Lindroos, D. Lunney, E. Mane, I. Podadera, Off-line commissioning of the ISOLDE cooler, Nuclear Instruments and Methods in Physics Research Section A, 266 (2008) 4502, https://doi.org/10.1016/j.nimb.2008.05.097.
[13] B.R. Barquest, J.C. Bale, J. Dilling, G. Gwinner, R. Kanungo, R. Krücke, M.R. Pearson, Development of a new RFQ beam cooler and buncher for the CANREB project at TRIUMF, Nuclear Instruments and Methods in Physics Research Section B, 376 (2016) 207, https://doi.org/10.1016/j.nimb.2016.02.035.
[14] T. Brunner, M.J. Smith, M .Brodeur, S. Ettenauer, A.T .Gallant, V.V. Simon, A. Chaudhuri, A. Lapierre, E. Mane´, R. Ringle, M.C .Simon, J.A. Vaz, P. Delheij, M. Good, M.R. Pearson, J. Dilling, TITAN's digital RFQ ion beam cooler and buncher, operation and performance, Nuclear Instruments and Methods in Physics Research Section A, 676 (2012) 32.
[15] A. A. Valverde, M. Brodeur, D. P. Burdette, J. A. Clark, J. W. Klimes, D. Lascar, P. D. O'Malley, R. Ringle, G. Savard, V. Varentsov, Stopped, bunched beams for the TwinSol facility, Hyperfine Interactions, (2019) 240:38, https://doi.org/10.1007/s10751-019-1591-x.
[16] A. Ruzzon, M. Maggiore, C. Roncolato, G. Ban, J.F. Cam, C. Gautier and C. Vandamme, Realization and simulations of the new SPES Beam Cooler, 2023 JINST 18 P10031, https://doi.org/10.1088/1748-0221/18/10/P10031.
[17] Yin-Shen Liu, Han-Rui Hu, Xiao-Fei Yang, Wen-Cong Mei, Yang-Fan Guo, Zhou Yan, Shao-Jie Chen, Shi-wei Bai, Shu-Jing Wang, Yong-Chao Liu, Peng Zhang, Dong-Yang Chen, Yan-Lin Ye, Qi-Te Li, Jie Yang, Stephan Malbrunot-Ettenauer, Simon Lechner, Carina Kanitz, Commissioning of a radiofrequency quadrupole cooler-buncher for collinear laser spectroscopy, Preprint arXiv:2502.10740v1 (physics), 15 Feb 2025, https://doi.org/10.48550/arXiv.2502.10740.
[18] M. Gerbaux, P. Ascher, A Husson, Antoine de Roubin, P. Alfaurt, M. Aouadi, Bertram Blank, Laurent Daudin, S El Abbeir, M Flayol, H. Guérin, S. Grevy, M. Hukkanen, B Lachacinski, David Lunney, S Perard, BARRY H. THOMAS, The General Purpose Ion Buncher: A radiofrequency quadrupole cooler-buncher for DESIR at SPIRAL2, Nuclear Instruments and Methods in Physics Research Section A 1046 (2023) 167631, https://doi.org/10.1016/j.nima.2022.167631.
[19] D.P. Burdette, R. Zite, M. Brodeur, A.A. Valverde, O. Bruce, R. Bualuan, A. Cannon, J.A. Clark, C. Davis, T. Florenzo, A.T. Gallant, J. Harkin, A.M. Houff, J. Li, B. Liu, J.Long, P.D. O'Malley, W.S.

Porter, C. Quick, R. Ringle, F. Rivero, G. Savard, M.A. Yeck, Off-line commissioning of the St. Benedict Radiofrequency Quadrupole Cooler-Buncher, Preprint arXiv:2504.08021v1, https://doi.org/10.48550/arXiv.2504.08021.

[20] V.A. Virtanen a, T. Eronen a, A. Kankainen a, O. Beliuskina a, P. Campbell b, R. Delaplanche c, Z. Ge a, R.P. de Groote d, M. Hukkanen a e, A. Jaries a, R. Kronholm a, I.D. Moore a, A. Raggio a, A. de Roubin f, J. Ruotsalainen a, M. Schuh, Miniaturised cooler-buncher for reduction of longitudinal emittance at IGISOL, Nuclear Instruments and Methods in Physics Research Section A 1072 (2025) 170186, https://doi.org/10.1016/j.nima.2024.170186.

[21] S. Lehner, S. SelsI. , Belosevic, F. Buchinger, P. Fischer, C. Kanitz, V.Lagaki, F.M. Maier, P. Plattner, L. Schweikhard, M. Vilen, S. Malbrunot-Ettenauer, Simulations of a cryogenic, buffer-gas filled Paul trap for low-emittance ion bunches, Nuclear Instruments and Methods in Physics Research Section A 1065 (2024) 169471, https://doi.org/10.1016/j.nima.2024.169471.

[22] Regan Zite, Maxime Brodeur, O. Bruce, D. Gan, P. D. O'Malley, William Samuel Porter, Fabio Rivero, Off-line Commissioning of the St. Benedict Radio Frequency Quadrupole Ion Guide, Nuclear Inst. and Methods in Physics Research, A 1088 (2026) 171483; https://doi.org/10.1016/j.nima.2026.171483.

[23] V. Varentsov, A New Approach to the Extraction System Design, SHIPTRAP Collaboration Meeting, Mainz, March 19, 2001; http://doi.org/10.13140/RG.2.2.30119.55200.

[24] Victor Varentsov, Review of Gas Dynamic RF-Only Funnel Technique for Low-Energy and High-Quality Ion Beam Extraction into a Vacuum, Micromachines 2023, 14(9), 1771; https://doi.org/10.3390/mi14091771.

[25] Victor Varentsov, Proposal of a New Double-Nozzle Technique for In-Gas-Jet Laser Resonance Ionization Spectroscopy, Atoms 2023, 11,88; https://doi.org/10.3390/atoms11060088.

[26] Victor Varentsov, The Double-Nozzle Technique Equipped with RF-Only Funnel and RF-Buncher for the Ion Beam Extraction into Vacuum, Atoms 2023, 11(10), 123; https://www.mdpi.com/2218-2004/11/10/123.

[27] Victor Varentsov, Towards a new and compact gas-dynamic cooler-buncher for the FAIR Laspec and MATS experiments, Journal of Instrumentation, , 20 (August 2025):P08031, https://doi.org/10.1088/1748-0221/20/08/P0803 1.

[28] Tim Ratajczyk, Philipp Bollinger, Tim Lellinger, Victor Varentsov, Wilfried Nörtershäuser, Towards a He-buffered laser ablation ion source for collinear laser spectroscopy, Hyperfine Interactions, (2020) 241:52, https://doi.org/10.1007/s10751-020-1698-0.

[29] Tim Ratajczyk, Kristian König, Philipp Bollinger, Tim Lellinger, Victor Varentsov, Wilfried Nörtershäuser and Julian Spahn, Transition frequencies, isotope shifts, and hyperfine structure in 4 s → 4 p transitions of stable Ti + ions, Physical Review A 100(3) 2024, http://dx.doi.org/10.1103/PhysRevA.110.032807.

[30] Tim Ratajczyk, Philipp Bollinger, Tim Lellinger, Victor Varentsov, Wilfried Nörtershäuser, A versatile laser ablation ion source for reference measurements at LASPEC, NUSTAR Collaboration Meeting, February 25 2021, http://doi.org/10.13140/RG.2.2.32743.11681.

[31] V.L. Varentsov. A.A. Ignatiev. Numerical investigations of internal supersonic jet targets formation for storage rings. Nuclear Instruments and Methods in Physics Research Section A 413 (1998) 447-456. http://dx.doi.org/10.1016/S0168-9002(98)00354-4.

[32] Scientific Instrument Services, Inc. 1027 Old York Road. Ringoes, NJ 08551-1054. USA. http://simion.com.

[33] V.L. Varentsov, N. Kuroda, Y. Nagata, H. A. Torii, M. Shibata, and Y. Yamazaki, ASACUSA Gas-Jet Target: Present Status And Future Development, AIP Conference Proceedings 793 (2005) 328-340, https://doi.org/10.1063/1.2121994

[34] D. Tiedemann, K.E. Stiebing, D.F.A. Winters, W. Quint, V. Varentsov, A. Warczak, A. Malarz, Th. Stöhlker, A pulsed supersonic gas jet target for precision spectroscopy at the HITRAP facility at GSI, Nucl. Instrum. Methods Phys. Res. A 764 (2014) 387–393, http://doi.org/10.1016/j.nima.2014.08.017